\documentclass[12pt]{article}
\usepackage[english]{babel}
\usepackage{amsmath}
\usepackage{hyperref}
\usepackage{amsfonts}
\usepackage{amssymb}
\begin{document}

\title{\bf{}BRST Lagrangian construction for spin-2 field in Einstein space.}

\author{\sc I.L. Buchbinder${}^{a}$\thanks{joseph@tspu.edu.ru},
V.A. Krykhtin${}^{ab}$\thanks{krykhtin@tspu.edu.ru}, P.M.
Lavrov$^c$\thanks{lavrov@tspu.edu.ru}
\\[0.5cm]
\it ${}^a$Department of Theoretical Physics,\\
\it Tomsk State Pedagogical University,\\
\it Tomsk 634061, Russia\\[0.3cm]
\it ${}^b$Laboratory of Mathematical Physics,\\
\it Tomsk Polytechnic University,\\
\it Tomsk 634034, Russia\\[0.3cm]
\it ${}^c$Department of Mathematical Analysis,\\
\it Tomsk State Pedagogical University,\\
\it Tomsk 634061, Russia}
\date{}

\maketitle
\thispagestyle{empty}

\begin{abstract}
We explore a new possibility of  BRST construction in higher spin
field theory to obtain a consistent Lagrangian for massive spin-2
field in Einstein space. Such approach automatically leads to
gauge invariant Lagrangian with suitable auxiliary and St\"uckelberg
fields. It is proved that in this case a propagation of spin-2 field
is hyperbolic and causal. Also we extend notion of partial
masslessness for spin-2 field in the background under consideration.
\end{abstract}

\section{Introduction}

Various aspects of higher spin field theory attract much
attention for a long time (see reviews \cite{reviews} and
references therein).
One of the modern approaches to deriving the Lagrangians for higher
spin fields is based on the use of BRST-BFV construction which was
initially developed for quantization of gauge theories \cite{BFV}.
Lagrangian formulation of the higher spin field theories within this
approach has been studied in \cite{brst1,brst2,brst3,brst4}.

At present, all Lagrangian formulations for higher spin fields,
including the BRST approach, are given for the background manifolds
corresponding to constant curvature spaces (Minkowski or AdS
spaces). Problem of finding the more general manifolds admitting the
consistent exact or approximate Lagrangian formulation for higher
spin fields is open in general. In BRST approach theory is
considered in terms of Fock vectors and the relations defining
on-shell contents are treated as the operator constraints acting on
the above vectors. Then, the problem of consistent Lagrangian
formulation appears as a problem of closing operator algebra
initiated by the constraints\footnote{We pay attention that not all
operators generating this algebra can be considered as the
constraints \cite{brst2,brst3}.} and related to both higher and
lower spin fields. In particular, the BRST approach in its existing
form, yields to Lagrangian formulation only for constant curvature
spaces even for spin-1 and spin-2 theories. However, it is evident
that spin-1 field can be formulated on the arbitrary curved
background. Moreover, it was shown some time ago that there exists a
consistent Lagrangian formulation for massive spin-2 field in
Einstein space\footnote{An Einstein space is defined by the relation
 $R_{\mu\nu}=const\cdot g_{\mu\nu}$ with Weyl tensor be arbitrary
(see e.g. \cite{1stein99fold}).}\cite{spin 2}.

In this note we want to pay attention to some new unexplored yet
possibility in the BRST approach to higher spin fields which allows us
to derive the known Lagrangian for massive spin-1 field without
restriction on the background and the Lagrangian for massive spin-2
field without restrictions on Weyl tensor. This possibility is
inherent just for BRST Lagrangian construction in higher spin theory
and has no counterpart in quantization theory\footnote{The other
differences of BRST approach to higher spin field theory from the
BRST approach to quantization of gauge theories are discussed in
\cite{brst2,brst3}.}.

The Lagrangian construction in the BRST approach
\cite{brst1,brst2,brst3,brst4}  was carried out for fields
of all spins simultaneously. The basic notions of this approach are
the Fock space vector $|\Psi_s\rangle$, corresponding to spin $s$
and the nilpotent BRST charge $Q$. The equations of motion and gauge
transformations are written in the form $Q|\Psi_s\rangle=0$ and
$\delta|\Psi_s\rangle=Q|\Lambda_s\rangle$ respectively, with the
BRST operator $Q$ being the same for all spins. Nilpotency of the
BRST operator provides us the gauge transformations and fields
$|\Psi_s\rangle$ and $|\Psi_s\rangle+Q|\Lambda_s\rangle$ are both
physical. Since we consider all spins simultaneously, then from
$Q^2|\Lambda_s\rangle=0$ follows $Q^2=0$. But if we want to
construct Lagrangian for the field with a given value $s$ of spin,
then it is sufficient to require a weaker condition that the BRST
operator for given spin $Q_s$ is not nilpotent in operator sense but
will be nilpotent only on the specific Fock vector parameter
$|\Lambda_s\rangle$ corresponding to a given spin $s$,
$Q_s^2|\Lambda_s\rangle=0$ only and $Q_s^2\neq0$ on states of
general form. This possibility has not been explored in all previous
applications of the BRST construction to the higher spin
fields\footnote{Such a possibility is inherent namely BRST approach
to higher spin field theory. In BFV-BRST approach to quantization of
gauge theories, unlike the higher spin field theory, one begins with
a given classical theory where BRST charge is constructed on the
base of given first class constraints and nilpotent by definition.}.
Just this point allows us to construct Lagrangians for spin-1 and
spin-2 field in background spaces different from Minkowski and AdS.

The paper is organized as follows. In the next section we describe
the BRST procedure of Lagrangian construction for the fields of fixed
spin on gravitational background. Then in sections \ref{secspin1}
and \ref{secspin2} we apply this method for Lagrangian construction
of fields with spin-1 and spin-2 respectively and in
section~\ref{causality} the causal propagation of spin-2 field is
proven. In section~\ref{summary} we summarize the results.

\section{General scheme of Lagrangian construction}

As is well known a completely symmetric bosonic field of spin-$s$
$\varphi_{\mu_1\ldots\mu_s}$
will realize irreducible representation of the Poincare group if the following equations are
satisfied
\begin{eqnarray}
(\partial^2-m^2)\varphi_{\mu_1\ldots\mu_s}=0,
\qquad
\partial^{\mu_1}\varphi_{\mu_1\ldots\mu_s}=0,
\qquad
\eta^{\mu_1\mu_2}\varphi_{\mu_1\mu_2\ldots\mu_s}=0.
\label{EM-flat}
\end{eqnarray}
When we turn to an arbitrary curved spacetime we suppose that conditions on
$\varphi_{\mu_1\ldots\mu_s}$ which must be satisfied,
tend to (\ref{EM-flat}) in the flat space limit.
It tells us that if we don't consider terms with the inverse powers of the
mass the equations on $\varphi_{\mu_1\ldots\mu_s}$
in curved spacetime must be of the form
\begin{eqnarray}
(\nabla^2-m^2)\varphi_{\mu_1\ldots\mu_s}
+\,\mbox{\it terms with curvature}=0,
\qquad
\nabla^{\mu_1}\varphi_{\mu_1\ldots\mu_s}=0,
\qquad
g^{\mu_1\mu_2}\varphi_{\mu_1\mu_2\ldots\mu_s}=0.
\label{EM-curved}
\end{eqnarray}
Our purpose is to find the ``terms with curvature''\footnote{Our definition of the curvature tensor is $R^{\alpha}_{\;\;\beta\mu\nu}=\partial_{\mu}\Gamma^{\alpha}_{\;\;\beta\nu}-
\partial_{\nu}\Gamma^{\alpha}_{\;\;\beta\mu}-
\Gamma^{\lambda}_{\;\;\beta\mu}\Gamma^{\alpha}_{\;\;\lambda\nu}+
\Gamma^{\lambda}_{\beta\nu}\Gamma^{\alpha}_{\;\;\lambda\mu}$} and
restriction on the curvature of the space demanding consistency
equations (\ref{EM-curved}) with each other and then to try to
construct Lagrangians for the higher spin fields in the background
gravity using the BRST method. Let us note that Lagrangian
construction can give additional restrictions on the spacetime
curvature apart from those which follow from consistency of
(\ref{EM-curved}).

To avoid explicit manipulations with a big number of indices it is
convenient to introduce the auxiliary Fock space generated by bosonic
creation and annihilation operators with tangent space indices
($a,b=0,1,\ldots,d-1$)
\begin{eqnarray}
[a_a,a^+_b]=\eta_{ab},
\qquad
\eta_{ab}=diag(-,+,\ldots,+).
\label{a-bos}
\end{eqnarray}
An arbitrary vector in this Fock space has the form
\begin{eqnarray}
\label{PhysState}
|\varphi\rangle
&=&
\sum_{s=0}^{\infty}\varphi_{a_1\ldots\,a_s}(x)\,
a^{+a_1}\ldots\,a^{+a_s}|0\rangle
=
\sum_{s=0}^{\infty}\varphi_{\mu_1\ldots\,\mu_s}(x)\,
a^{+\mu_1}\ldots\,a^{+\mu_s}|0\rangle
\equiv
\sum_{s=0}^{\infty}|\varphi_s\rangle
,
\end{eqnarray}
where
$a^{+\mu}(x)=e^\mu_a(x)a^{+a}$, $a^\mu(x)=e^\mu_a(x)a^a$,
with $e^\mu_a(x)$ being the vielbein.
It is evident that $[a_\mu, a_\nu^+]=g_{\mu\nu}$.
We also
suppose the standard relation$\nabla_\mu e^a_\nu=0$.

Then one
introduces derivative operator
\begin{eqnarray}
D_\mu=\partial_\mu+\omega_\mu^{ab}a_a^+a_b, &\qquad& D_\mu|0\rangle=0
\label{D_mu}
\end{eqnarray}
and realize equations (\ref{EM-curved}) in the operator
form
\begin{eqnarray}
l_0|\varphi_s\rangle=
l_1|\varphi_s\rangle=
l_2|\varphi_s\rangle=
0
\label{EM-op}
\end{eqnarray}
where
\begin{eqnarray}
\label{l012}
&&
l_0=D^2-m^2+\mathcal{X},
\qquad
l_1=-ia^\mu D_\mu,
\qquad
l_2={\textstyle\frac{1}{2}}\,a^\mu a_\mu,
\\
&&
\qquad
D^2=g^{\mu\nu}(D_\mu D_\nu-\Gamma_{\mu\nu}^{\sigma}D_\sigma)
\end{eqnarray}
with the operator $\mathcal{X}$ corresponding to the ``terms with curvature''
in (\ref{EM-curved}).

Since operators $l_1$ and $l_2$ commute $[l_1,l_2]=0$ the
consistency of equations (\ref{EM-curved}) demands
\begin{eqnarray}
[l_0,l_1]|\varphi_s\rangle\sim0,
&\qquad&
[l_0,l_2]|\varphi_s\rangle\sim0,
\label{eqX}
\end{eqnarray}
where $\sim$ means ``up to equations of motion (\ref{EM-curved}) is equal to''.
Equations (\ref{eqX}) are the equations on operator $\cal{X}$ and on the background
gravity.

Having found operator $\cal{X}$ and the restrictions on the spacetime
curvature from (\ref{eqX}) we will try to construct Lagrangians for the field with given spin using the BRST method.
The procedure of Lagrangian construction is as follows.
For the Lagrangian be a real function the BRST operator used for its construction must
be a Hermitian operator. It assumes that the set of operators underlying the BRST
operator must be invariant under Hermitian conjugation.
Thus to have such a set of operators we add to constraints $l_0$, $l_1$, $l_2$ their
Hermitian conjugated operators. Since $l_0$ is assumed to be self conjugated we
add two operators
\begin{eqnarray}
l_1^+=-ia^{\mu+}D_\mu
&\qquad&
l_2^+=\tfrac{1}{2}a^+_\mu a^{\mu+}
.
\label{l12+}
\end{eqnarray}
Then for constructing the BRST operators the underlying set of
operators must form an algebra. Note that the nilpotency condition
of the BRST operators is needed for existing of gauge symmetry. As it
is known if we consider spin-s field and decompose the gauge parameter
$|\Lambda_s\rangle$ in series of creation operators,
then maximal tensorial rank of gauge parameters
$|\lambda_{k}\rangle=a^{+\mu_1}\cdots{}a^{+\mu_{k}}\lambda_{\mu_1\cdots\mu_{k}}|0\rangle$,
entering in $|\Lambda_s\rangle$ is $k=s-1$ (see e.g. \cite{brst2,brst3}). Therefore
if we want to construct Lagrangian for a particular spin-s field it
is enough that this set of operators form algebra only on states
$|\lambda_{k}\rangle$
 with $k<s$. To
form an algebra we must add to $l_0$, $l_1$, $l_2$, $l_1^+$, $l_2^+$
all the operators generated by the commutators of these operators.
But if we want to construct with the help of the obtained algebra
Lagrangian for spin-s field, then this algebra must be a deformation
of the algebra in Minkowski \cite{brst2} (or in AdS \cite{brst3})
space. Thus we can add only two operators which are generalization
of operators
\begin{eqnarray}
g_0=a_\mu^+a^\mu+\tfrac{d}{2}
\qquad
g_m=m^2+const
\label{g0}
\end{eqnarray}
to the case of curved space. Since operator $g_0$ is dimensionless
and we do not consider terms with inverse power of the mass then it
is impossible to deform operator $g_0$ by terms with curvature.
Therefore operator $g_0$ keeps the same form (\ref{g0}) as in the
flat case. Let us turn to the operator $g_m$. If we add to it any
nonconstant term then the algebra will not be closed since in this
case (for example) commutator $[l_1,g_m]$ will not be proportional
to the operators of the algebra. Again the form of operators $g_m$
(\ref{g0}) must be kept. Thus we came to the conclusion that in
order to construct Lagrangian with the help of the BRST method we
must find operators $l_0$ and $g_m$ so that operators $l_0$, $l_1$,
$l_2$, $l_1^+$, $l_2^+$, $g_0$, $g_m$ form an algebra on states
$|\lambda_{k}\rangle$ with $k<s$. One should note that the obtained
algebra must satisfy Jacobi identity. After the algebra is
constructed the Lagrangian construction procedure is the same as in
flat or AdS case. It is unclear how the above requirements may be
fulfilled in spaces with arbitrary curvature for any spin-s. As we
show later using this method it is possible to construct Lagrangian
for spin-1 field in arbitrary curved space and for spin-2 case in Einstein space.
Let us consider application of the above method.

\section{BRST Lagrangian construction for spin-1 field}\label{secspin1}

Let us first consider spin-1 field as an example of the above described procedure of
Lagrangian construction.
In this case we will be looking for operator $\mathcal{X}$ in the form
\begin{eqnarray}
\mathcal{X}=X(x)+X^{\mu\nu}(x)a^+_\mu a_\nu
\end{eqnarray}
where yet unknown functions $X(x)$ and $X^{\mu\nu}(x)$ should be
defined from the consistency condition (\ref{eqX}). Since the
tracelessness condition for the spin-1 case is redundant therefore
it is sufficient to consider only the following commutator
\begin{eqnarray}
[l_0,l_1]|\varphi_1\rangle
&=&
i(R^\nu_\mu+X^\nu{}_\mu)a^\mu D_\nu|\varphi_1\rangle
+i(\tfrac{1}{2}R_{;\mu}+X_{;\mu}+X^\nu{}_{\mu;\nu})a^\mu|\varphi_1\rangle
.
\end{eqnarray}
From the first summand in the right hand side one finds
\begin{eqnarray}
X^{\mu\nu}=-R^{\mu\nu}+F(x)g^{\mu\nu},
\end{eqnarray}
where $F(x)$ is an arbitrary scalar function with the dimension of mass
squared.
Substituting this expression for $X^{\mu\nu}$ into the second summand of the
right hand side one gets
\begin{eqnarray}
X=-F(x)
\end{eqnarray}
ignoring a possible constant with the dimension of mass squared
which can be absorbed by redefinition of the mass term. Collecting
together one gets a partial solution to operator $l_0$ in the form
\begin{eqnarray}
\tilde{l}_0=D^2-m^2-R^{\mu\nu}a^+_\mu a_\nu+F(x)(a^+_\mu a^\mu-1).
\end{eqnarray}
Since $(a^+_\mu a^\mu-1)|\varphi_1\rangle=0$ the term proportional to $F(x)$
doesn't influence on the mass-shell equation in the component form.
In what follows we put $F(x)=0$ and take operator $l_0$ in the form
\begin{eqnarray}
l_0=D^2-m^2-R^{\mu\nu}a^+_\mu a_\nu.
\end{eqnarray}

Let us turn to Lagrangian construction. To get a set of operators invariant under Hermitian conjugation we add to constraints $l_0$, $l_1$ one more operator
\begin{eqnarray}
 l_1^+=-ia^{+\mu}D_\mu
\end{eqnarray}
and close the set of operators $l_0$, $l_1$, $l_1^+$ to an algebra adding one more operator
\begin{eqnarray}
g_m=m^2.
\end{eqnarray}
The algebra has only one non-vanishing commutator
\begin{eqnarray}
[l_1,l_1^+]=-l_0-g_m.
\label{1-l1l1+}
\end{eqnarray}
Let us remind that commutator (\ref{1-l1l1+}) is understood as
$[l_1,l_1^+]|\lambda_0\rangle=(-l_0-g_m)|\lambda_0\rangle$. It is
evident that Jacobi identities are satisfied. Thus, the algebra is
obtained and further Lagrangian construction goes in the usual way
\cite{brst2,brst3}.

Since in the set of operators of the algebra there is operator $g_m$
which is not a constraint neither in the space of ket-vectors nor in
the space of bra-vectors one should construct new enlarged
expressions for the operators $L_i=l_i+l_i'$, $G_m=g_m+g_m'$ so that
$G_m=0$ with the same algebra as for the initial operators $l_i$,
$g_m$ (see \cite{brst3} for more details). Introducing new pair of
bosonic creation and annihilation operators $b^+$, $b$ with the
standard commutation relations $[b,b^+]=1$ we define the enlarged
expressions for the operators as follows
\begin{eqnarray}
L_0=l_0
\qquad
L_1=l_1+mb
\qquad
L_1^+=l_1^++mb^+
\qquad
G_m=0
\end{eqnarray}
and then construct the BRST operator
\begin{eqnarray}
&&
Q=\eta_0l_0+\eta_1^+L_1+\eta_1L_1^++\eta_1^+\eta_1\mathcal{P}_0
\end{eqnarray}
with fermionic ghosts
\begin{eqnarray}
\{\eta_1,\mathcal{P}_1^+\}=1,
\qquad
\{\eta_1^+,\mathcal{P}_1\}=1,
\qquad
\{\eta_0,\mathcal{P}_0\}=1
\end{eqnarray}
which acts on the vacuum state as
\begin{eqnarray}
\eta_1|0\rangle=\mathcal{P}_1|0\rangle=\mathcal{P}_0|0\rangle=0.
\end{eqnarray}

Finally, Lagrangian (up to an overall factor) and gauge
transformations in concise notation are (see \cite{brst2,brst3} for details)
\begin{eqnarray}
\mathcal{L}&=&\int d\eta_0\langle\Psi_1|Q|\Psi_1\rangle,
\qquad
\delta|\Psi_1\rangle=Q|\Lambda_1\rangle,
\end{eqnarray}
where
\begin{eqnarray}
&&
|\Psi_1\rangle=\Bigl\{-ia^{+\mu}A_\mu(x)+b^+A(x)+\eta_0\mathcal{P}_1^+\varphi(x)\Bigr\}|0\rangle,
\qquad
|\Lambda_1\rangle=\mathcal{P}_1^+\lambda(x)|0\rangle.
\end{eqnarray}
In component form the Lagrangian and gauge transformations are
\begin{eqnarray}
\mathcal{L}&=&
\frac{1}{2}A^\mu\Bigl\{(\nabla^2-m^2)A_\mu-R_{\mu\nu}A^\nu-\nabla_\mu\varphi\Bigr\}
+\frac{1}{2}A\Bigl\{(\nabla^2-m^2)A-m\varphi\Bigr\}
\\
&&{}
+\frac{1}{2}\varphi\Bigl\{\nabla^\mu A_\mu-mA-\varphi\Bigr\}
\\
&&
\delta A_\mu=\nabla_\mu\lambda,
\qquad
\delta A=m\lambda,
\qquad
\delta\varphi=(\nabla^2-m^2)\lambda.
\end{eqnarray}
After elimination of the auxiliary fields the Lagrangian takes its usual form
\begin{eqnarray}
\mathcal{L}&=&-\frac{1}{4}F_{\mu\nu}F^{\mu\nu}-\frac{m^2}{2}A_\mu A^\mu
\end{eqnarray}
where $F_{\mu\nu}=\nabla_\mu A_\nu-\nabla_\nu A_\mu$. Thus we have
reproduced the Lagrangian for massive spin one field in arbitrary
curved space using the BRST method.

\section{Lagrangian construction for spin-2 field}\label{secspin2}
In spin-2 case
we will search for a partial solution to (\ref{eqX}) for operator $\mathcal{X}$
in the form
\begin{eqnarray}
\mathcal{X}=X(x)+X^{\mu\nu}(x)a^+_\mu a_\nu
+X^{\mu\nu\alpha\beta}a^+_\mu a^+_\nu a_\alpha a_\beta
.
\label{X-2}
\end{eqnarray}
To find coefficients in (\ref{X-2}) we consider commutator
\begin{eqnarray}
[l_0,l_1]|\varphi_2\rangle&=&
2i(R^{\mu\alpha\beta\nu}+X^{\mu\nu\alpha\beta})
a^+_\mu a_\alpha a_\beta D_\nu|\varphi_2\rangle
+i(R^{\sigma\alpha}+X^{\sigma\alpha})a_\alpha D_\sigma|\varphi_2\rangle
\nonumber
\\
&&{}
+i(X_{\mu\alpha;\beta}+2X_{\mu\sigma\alpha\beta}{}^{;\sigma}+R_{\mu\alpha;\beta}-R_{\alpha\beta;\mu})
a^{\mu+}a^\alpha a^\beta|\varphi_2\rangle
\nonumber
\\
&&{}
+i(X_{;\alpha}+X^\sigma{}_{\alpha;\sigma}+\tfrac{1}{2}R_{;\alpha})a^\alpha |\varphi_2\rangle
.
\label{2-l0l1}
\end{eqnarray}
From the first two summand of the right hand side of (\ref{2-l0l1}) and the
last one we find
\begin{eqnarray}
&&
X_{\mu\nu\alpha\beta}=-R_{\mu(\alpha\beta)\nu}
+F_1(x)g_{\mu\nu}g_{\alpha\beta}
+F_2(x)(g_{\mu\alpha}g_{\nu\beta}+g_{\mu\beta}g_{\alpha\nu})
\\
&&
X_{\mu\nu}=-R_{\mu\nu}+F_3(x)g_{\mu\nu}
\\
&&
X=-F_3(x)+C
\end{eqnarray}
where $F_i(x)$ are arbitrary functions with dimension of mass squared and $C$
is an arbitrary constant with the same dimension.
Let us substitute the found coefficients of the operator $\mathcal{X}$
(\ref{X-2}) into the third summand of the right hand side of (\ref{2-l0l1})
\begin{eqnarray}
[l_0,l_1]|\varphi_2\rangle&=&
(4F_2+F_3)_{;\alpha}g_{\mu\beta}a^{+\mu}a^\alpha a^\beta|\varphi_2\rangle
+(R_{\alpha\beta;\mu}-2R_{\mu(\alpha;\beta)}) a^{+\mu}a^\alpha a^\beta|\varphi_2\rangle
.
\end{eqnarray}
To provide (\ref{eqX}) we have to suppose that $\nabla_\alpha(4F_2+F_3)=0$
and
\begin{eqnarray}
R_{\alpha\beta;\mu}=2R_{\mu(\alpha;\beta)}
\;\Rightarrow\;
R=const
.
\label{2-R}
\end{eqnarray}
We see that consistent equations of motion for spin-2 field exist only in
space with Ricci and scalar curvature satisfying (\ref{2-R}) while Weyl
tensor are arbitrary.
As a result a partial solution to (\ref{eqX}) has the form
\begin{eqnarray}
\tilde{l}_0&=&D^2-m^2-R^{\mu\alpha\beta\nu}a_\mu^+ a_\nu^+ a_\alpha a_\beta
-R^{\mu\nu}a_\mu^+ a_\nu
\nonumber
\\
&&{}
+4F_1l_2^+l_2+2F_2(N-1)(N-2)+\xi_1RN+\xi_2R,
\qquad
N=a_\mu^+a_\mu
\end{eqnarray}
where $F_1$ and $F_2$ are arbitrary functions with dimension of mass squared
and $\xi_1$ and $\xi_2$ are arbitrary dimensionless constants.
It is easy to check that $[\tilde{l}_0,l_2]|\varphi_2\rangle\sim0$.
Now for future convenience we put $F_1=F_2=0$ and constants $\xi_1$ and $\xi_2$ we will fix later.

Let us turn to Lagrangian construction using the BRST method. First,
in accordance with this method \cite{brst2,brst3} we should
add to operators $l_0$, $l_1$, $l_2$ their Hermitian conjugated
operators. Since $l_0$ is Hermitian operator we add only $l_1^+$ and
$l_2^+$
\begin{eqnarray}
l_1^+=-ia^{+\mu}D_\mu
\qquad
l_2^+=\tfrac{1}{2}a^{+\mu}a_\mu^+.
\end{eqnarray}
Next we must add to
$l_0$, $l_1$, $l_2$, $l_1^+$, $l_2^+$
two more operators $g_0$, $g_m$ (\ref{g0}) to close the algebra, with the ``$const$'' in $g_m$ to be defined.
To define it we consider commutators $[l_1,l_1^+]$ acting  on states $|\lambda_k\rangle$ with $k<2$
\begin{eqnarray}
[l_1,l_1^+]
&=&-D^2+R_{\mu\nu\alpha\beta}a^{+\mu}a^\nu a^{+\alpha}a^\beta
\nonumber
\\
&=&
-\tilde{l}_0-m^2+(1-\tfrac{d}{2}\xi_1+\xi_2)R
+(\xi_1-\tfrac{2}{d})Rg_0
\nonumber
\\
&&{}
-2\tilde{R}^{\mu\nu}a_\mu^+a_\nu
-2R^{\mu\alpha\beta\nu}a_\mu^+a_\nu^+a_\alpha a_\beta
,
\label{tilde}
\end{eqnarray}
where $\tilde{R}_{\mu\nu}=R_{\mu\nu}-\frac{1}{d}g_{\mu\nu}R$ is the traceless part
of the Ricci tensor.
Since it is supposed that (\ref{tilde}) acts on states $|\lambda_k\rangle$ with $k<2$
and since $R=const$ then the right hand side of (\ref{tilde}) is expressed through
the operators of the algebra only in the case when $\tilde{R}_{\mu\nu}=0$.
Now we fix the arbitrary constants $\xi_1$, $\xi_2$ so that the mass-shell operators
(which we denote $l_0$ instead of $\tilde{l_0}$) becomes
\begin{eqnarray}
&&
l_0=D^2-m^2-R^{\mu\alpha\beta\nu} a_\mu^+a_\nu^+a_\alpha a_\beta
-\tfrac{1}{d}R(g_0-\tfrac{d}{2}-2)
.
\label{l0}
\end{eqnarray}
Thus commutator $[l_1,l_1^+]$ takes the form
\begin{eqnarray}
[l_1,l_1^+]
&=&
-l_0-m^2+\tfrac{d+2}{d}R-\tfrac{2}{d}Rg_0
-2R^{\mu\alpha\beta\nu}a_\mu^+a_\nu^+a_\alpha a_\beta
.
\end{eqnarray}
Of course, if we are supposing that the commutator acts on
$|\lambda_{k<2}\rangle$ then the last term with Riemann tensor
containing two annihilation operators may be discarded completely.
But this leads to violation of Jacobi identities. One can check that
if we discard only part of the last term with the Weyl tensor
$C^{\mu\alpha\beta\nu}a_\mu^+a_\nu^+a_\alpha a_\beta$ then the
Jacobi identities will be satisfied. As a result commutator
$[l_1,l_1^+]$ should be taken in the form
\begin{eqnarray}
[l_1,l_1^+]&=&
-l_0-g_m-\tfrac{2}{d(d-1)} \,R\, [g_0^2-2g_0-4l_2^+l_2]
\label{l1l1+}
\end{eqnarray}
with
\begin{eqnarray}
g_m=m^2-\tfrac{d^2-4}{2d(d-1)}R
\end{eqnarray}
Thus algebra of the operators
$l_0$, $l_1$, $l_1^+$, $l_2$, $l_2^+$, $g_0$, $g_m$
has the form given in Table~\ref{table}.
\begin{table}
\begin{eqnarray*}
&
\begin{array}{||l||r|r|r|r|r||r|c||}\hline\hline
\left[\; \downarrow, \rightarrow \right]
\vphantom{\biggm|}
       &l_0&l_1&l_1^+&l_2&l_2^+&g_0&\quad g_m\quad\\
\hline\hline\vphantom{\biggm|}
l_0
          &0&0&0&0&0&0&0
              \\
\hline\vphantom{\biggm|}
l_1
          &0&0&(\ref{l1l1+})&0&l_1^+&l_1&0 \\
\hline\vphantom{\biggm|}
l_1^+
          &0&-(\ref{l1l1+})&0&-l_1&0&-l_1^+&0 \\
\hline\vphantom{\biggm|}
l_{2}
          &0&0&l_1&0&g_0&2l_2&0\\
\hline\vphantom{\biggm|}
l_{2}^+
          &0&-l_1^+&0&-g_0&0&-2l_2^+&0\\
\hline\hline\vphantom{\biggm|}
g_{0}
          &0&-l_1&l_1^+&-2l_2&2l_2^+&0&0\\
\hline\vphantom{\biggm|}
g_m
          &0&0&0&0&0&0&0\\
\hline\hline
\end{array}
\end{eqnarray*}
\caption{Algebra of the operators}
\label{table}
\end{table}
In this Table
the first arguments of the commutators are in
the first column, the second arguments are in the upper row.
One can check that algebra given in Table~\ref{table} satisfies Jacobi identities.

Now the procedure of Lagrangian construction is the same as in
\cite{brst3} and we will give only its main steps without details.
First, since we have two operators $g_0$ and $g_m$ which are not
constraints neither in the bra-vector nor in the ket-vector spaces
one should construct extended expressions for the operators
$l_i\to{}L_i=l_i+l_i'$ in order for these operators not to give
supplementary equations on the physical state. Introducing two pairs
of bosonic creation and annihilation operators $b_1^+$, $b_1$ and
$b_2^+$, $b_2$ with the standard commutation relations
$[b_1,b_1^+]=[b_2,b_2^+]=1$ the result for the additional parts
$l_i'$ has the form
\begin{align}
&
l_0'=0
&&
l_1^{\prime+}=m_1b_1^+
&&
l_2^{\prime+}=b_2^+
&&
g_0'=b_1^+b_1+b_2^+b_2-\tfrac{d-2}{2}
&&
g_m'=\tfrac{d^2-4}{2d(d-1)}R-m^2
\end{align}
\begin{eqnarray}
l_1'=m_1b_1^+b_2
+\tfrac{m^2}{m_1}\,b_1
-\tfrac{R}{m_1d}\,b_1^+b_1^2
+\ldots
&\qquad&
l_2'=-\tfrac{d-2}{2}\,b_2
+\tfrac{m^2}{2m_1}\,b_1^2
+\ldots
\end{eqnarray}
where $m_1$ is an arbitrary constant parameter with dimension of
mass which is constructed from $m$ and $R$, $m_1=f(m,R)\neq0$ and
the dots stand for the terms irrelevant for spin-2 case.

Then we construct BRST operator on the base of the extended
operators $L_i$ and extract from it the part which is used for
Lagrangian construction\footnote{$Q_2$ in (\ref{Q2}) is analog of
operator $Q_s$ in \cite{brst3} formula (80).}
\begin{eqnarray}
Q_2&=&
\eta_0L_0+\eta_1^+L_1+\eta_1L_1^++\eta_2^+L_2+\eta_2L_2^+
+\eta_1^+\eta_1\mathcal{P}_0
-\eta_1^+\eta_2\mathcal{P}_1^+
-\eta_2^+\eta_1\mathcal{P}_1
\nonumber
\\
&&{}
+\frac{4R}{d(d-1)}\eta_1^+\eta_1\Bigl\{
(L_2^+-2l_2^{\prime+})\mathcal{P}_2+(L_2-2l_2')\mathcal{P}_2^+\Bigr\}
\label{Q2}
\end{eqnarray}
with the ghosts satisfying the anticommutation relations
\begin{eqnarray}
\{\eta_1,\mathcal{P}_1^+\}=1,
\quad
\{\eta_1^+,\mathcal{P}_1\}=1,
\quad
\{\eta_2,\mathcal{P}_2^+\}=1,
\quad
\{\eta_2^+,\mathcal{P}_2\}=1,
\quad
\{\eta_0,\mathcal{P}_0\}=1.
\end{eqnarray}

Now Lagrangian and gauge transformations in terms of operator $Q_2$ (\ref{Q2}) are
written as follows
\begin{eqnarray}
\label{L2}
\mathcal{L}&=&\int d\eta_0\langle\Psi_2|KQ_2|\Psi_2\rangle
\qquad
\delta|\Psi_2\rangle=Q_2|\Lambda_2\rangle
\end{eqnarray}
where operator $K$
\begin{eqnarray}
K&=&|0\rangle \langle0|
+\frac{m^2}{m_1^2}b_1^+|0\rangle \langle0|b_1
+\frac{m^2}{2m_1^4}\Bigl(m^2-\frac{R}{d}\Bigr)b_1^{+2}|0\rangle \langle0|b_1^2
\nonumber
\\
&&{}
-\frac{d-2}{2}\,b_2^+|0\rangle \langle0|b_2
+\frac{m^2}{2m_1^2}
\Bigl(b_1^{+2}|0\rangle \langle0|b_2+b_2^+|0\rangle \langle0|b_1^2\Bigr)
+\ldots
\end{eqnarray}
provides the reality of the Lagrangian and states $|\Psi_2\rangle$ and
$|\Lambda_2\rangle$ are
\begin{eqnarray}
|\Psi_2\rangle&=&|\Phi_2\rangle+\eta_1^+\mathcal{P}_1^+|\Phi_0\rangle
+\eta_0\mathcal{P}_1^+|\Phi_1\rangle
+\eta_0\mathcal{P}_2^+|\Phi_0'\rangle
\qquad
|\Lambda_2\rangle=\mathcal{P}_1^+|\lambda_1\rangle+\mathcal{P}_2^+|\lambda_0\rangle
\\
&&
|\Phi_2\rangle= \Bigr[
{\textstyle\frac{(-i)^2}{2}}\,a^{\mu+}a^{\nu+}H_{\mu\nu}(x)
+b_2^+H_1(x) -ia^{\mu+}b_1^+A_\mu(x) +b_1^{+2}\varphi(x)
\Bigl]|0\rangle
\\
&&
|\Phi_0\rangle=H(x)|0\rangle,
\hspace*{3ex}
|\Phi_1\rangle= \Bigl[ -ia^{\mu+}H_\mu(x)+b_1^+A(x) \Bigr]|0\rangle,
\hspace{3ex}
|\Phi_0'\rangle=H_2(x)|0\rangle,
\\
&&
|\lambda_1\rangle= \Bigl[ -ia^{\mu+}\lambda_\mu(x)+b_1^+\lambda(x)
\Bigr]|0\rangle,
\hspace{3ex}
|\lambda_0\rangle=\lambda_2(x)|0\rangle.
\end{eqnarray}
From (\ref{L2}) one can find Lagrangian in component form
\begin{eqnarray}
\mathcal{L}
&=&
\tfrac{1}{2}H^{\mu\nu}\Bigl\{
(\nabla^2-m^2)H_{\mu\nu}-2R_{\mu\alpha\beta\nu}H^{\alpha\beta}
-2\nabla_\mu H_\nu+g_{\mu\nu}H_2
\Bigr\}
\nonumber
\\
&&{}
+\tfrac{m^2}{m_1^2}A^\mu\Bigl\{
(\nabla^2-m^2+\tfrac{R}{d})A_\mu
-m_1H_\mu-\nabla_\mu A
\Bigr\}
\nonumber
\\
&&{}
+\tfrac{m^2}{m_1^2}\Bigl[H_1+2\varphi\Bigl(\tfrac{m^2}{m_1^2}-\tfrac{R}{dm_1^2}\Bigr)\Bigr]\Bigl\{
(\nabla^2-m^2+\tfrac{2R}{d})\varphi-m_1A
\Bigr\}
\nonumber
\\
&&{}
+\Bigl(\tfrac{m^2}{m_1^2}\varphi-\tfrac{d-2}{2}H_1\Bigr)\Bigl\{
(\nabla^2-m^2+\tfrac{2R}{d})H_1-H_2
\Bigr\}
\nonumber
\\
&&{}
-H\Bigl\{
(\nabla^2-m^2+\tfrac{2R}{d})H+\nabla^\mu H_\mu-\tfrac{m^2}{m_1}A-H_2
\Bigr\}
\nonumber
\\
&&{}
-H^\mu\Bigl\{
-\nabla^\nu H_{\mu\nu}+\tfrac{m^2}{m_1}A_\mu-\nabla_\mu H+H_\mu
\Bigr\}
\nonumber
\\
&&{}
-\tfrac{m^2}{m_1^2}A\Bigl\{
-\nabla^\mu A_\mu+m_1H_1
+\tfrac{2m^2}{m_1}\varphi-\tfrac{2R}{dm_1}\varphi-m_1H+A
\Bigr\}
\nonumber
\\
&&{}
-H_2\Bigl\{
-\tfrac{1}{2}H^\mu_\mu-\tfrac{d-2}{2}H_1+\tfrac{m^2}{m_1^2}\varphi-H
\Bigr\}
\label{L2c}
\end{eqnarray}
and gauge transformations
\begin{align}
&
\delta{}H_{\mu\nu}
=
\nabla_\mu\lambda_\nu+\nabla_\nu\lambda_\mu
-g_{\mu\nu}\lambda_2
,
&&
\delta{}H_1=\lambda_2,
\\
&
\delta{}A_\mu=\nabla_\mu\lambda+m_1\lambda_\mu,
&&
\delta\varphi=m_1\lambda,
\\
&
\delta{}H=
-\nabla^\mu\lambda_\mu+\frac{m^2}{m_1}\,\lambda
+\lambda_2,
&&
\delta{}H_\mu
=
(\nabla^2-m^2+\tfrac{R}{d})\lambda_\mu
\\
&
\delta{}A
=
(\nabla^2-m^2+\tfrac{2R}{d})\lambda
&&
\delta{}H_2
=
(\nabla^2-m^2+\tfrac{2R}{d})\lambda_2
.
\end{align}
Thus we have derived Lagrangian for spin-2 field in the Einstein spacetime.

Now we can remove a part or all the auxiliary fields. Doing the
transformations analogous to \cite{brst3} one obtains the following
Lagrangian
\begin{eqnarray}
\mathcal{L}
&=&
\tfrac{1}{2}H^{\mu\nu}\Bigl\{
(\nabla^2-m^2-\tfrac{2R}{d(d-1)})H_{\mu\nu}-2\nabla_\mu\nabla^\sigma H_{\nu\sigma}
-2C_{\mu\alpha\beta\nu}H^{\alpha\beta}
+2\nabla_\mu\nabla_\nu H
\Bigr\}
\nonumber
\\
&&{}
+\tfrac{m^2}{m_1^2}A^\mu\Bigl\{
(\nabla^2+\tfrac{R}{d})A_\mu
-\nabla_\mu\nabla^\nu A_\nu
-2m_1\nabla^\nu H_{\mu\nu}+2m_1\nabla_\mu H
\Bigr\}
\nonumber
\\
&&{}
+\tfrac{m^2}{m_1^2}2\varphi\Bigl(\tfrac{d-1}{d-2}\tfrac{m^2}{m_1^2}-\tfrac{R}{dm_1^2}\Bigr)\Bigl\{
(\nabla^2+\tfrac{dm^2}{d-2})\varphi-2m_1\nabla^\mu A_\mu+m_1^2 H
\Bigr\}
\nonumber
\\
&&{}
-\tfrac{1}{2}H\Bigl(\nabla^2-m^2+\tfrac{R(d-3)}{d(d-1)}\Bigr)H
\label{L2-Zin}
\end{eqnarray}
\begin{align}
&
\delta{}H_{\mu\nu}
=
\nabla_\mu\lambda_\nu+\nabla_\nu\lambda_\mu
-g_{\mu\nu}\lambda\tfrac{2m^2}{(d-2)m_1}
,
&&
\delta{}A_\mu=\nabla_\mu\lambda+m_1\lambda_\mu,
&&
\delta\varphi=m_1\lambda,
\end{align}
with $H=H^\mu_\mu$. The Lagrangian (\ref{L2-Zin}) contains Weyl
tensor $C_{\mu\alpha\beta\nu}$ in explicit form. If the
$C_{\mu\alpha\beta\nu}=0$ one gets, after some field redefinition,
the Lagrangian given in \cite{Zinoviev} for the constant curvature
background space.

It is easy to see that if $m^2=\frac{d-2}{d(d-1)}R$, then
scalar field $\varphi$ disappears from Lagrangian (\ref{L2-Zin}). As
a result such a Lagrangian in $d=4$ describes propagation of the helicities
$\pm2$, $\pm1$, thus the field $H_{\mu\nu}$ becomes partial massless
in terminology \cite{Deser}, although in (\ref{L2-Zin}) we did not
assume that Weyl tensor vanished.

Finally one can write Lagrangian in terms of physical field alone
\begin{eqnarray}
\mathcal{L}
&=&
\tfrac{1}{2}H^{\mu\nu}\Bigl\{
(\nabla^2-m^2+\tfrac{2R}{d})H_{\mu\nu}-2\nabla^\sigma\nabla_\mu H_{\nu\sigma}
+2\nabla_\mu\nabla_\nu H
\Bigr\}
\nonumber
\\
&&{}
-\tfrac{1}{2}H(\nabla^2-m^2+\tfrac{R}{d})H
\label{L2-FP}
\end{eqnarray}
with no gauge symmetry.

\section{Causality of massive spin-2 field propagation}\label{causality}

Now we turn to the problem of causality of massive spin-2 field with
Lagrangian (\ref{L2-FP}) in Einstein space. It is well known that
the higher spin field theory faces a problem of inconsistency, in
particular, coupling to external field can violate a causality of
free theory\footnote{Aspects of consistency and causality for spin-2
field in external background are discussed in \cite{Velo},
\cite{spin 2}, \cite{Waldron}.}.
Our consideration is based on the method of Velo and Zwanziger
\cite{Velo} adapted to the theories in curved spacetime.

We begin with a brief outline of the method.  If one has a system of the
second order differential equations for a set of fields $\varphi^B$
\begin{eqnarray}
M^A_{B}{}^{\mu\nu}\partial_\mu\partial_\nu\varphi^B+\ldots=0,
\qquad
\mu,\nu=0,\ldots,{}d-1
\label{System}
\end{eqnarray}
then to verify that the system (\ref{System}) describes hyperbolic propagation
one should check that all solutions $n_0(n_i)$, $(i=1,\ldots{}d-1)$  of the algebraic equation
\begin{eqnarray}
\det(M^A_B{}^{\mu\nu}n_\mu n_\nu)=0
\label{Det}
\end{eqnarray}
are real for any given real set of $n_i$.
The hyperbolic system is called causal if there are no timelike vectors among
solutions $n_\mu$ of (\ref{Det}).

In many physical cases (including spin-2 field) equation (\ref{Det})
fulfils identically. In this case one should replace the initial
system of equations by another equivalent system of equations
supplemented by constraints at a given initial time. Then the above
analysis should be applied to this new system.

Let us turn to our spin-2 field described by Lagrangian (\ref{L2-FP}).
The equations of motion are
\begin{eqnarray}
E_{\mu\nu}&\equiv&
(\nabla^2-m^2+\tfrac{2R}{d})H_{\mu\nu}
-2\nabla^\sigma\nabla_{(\mu} H_{\nu)\sigma}
+\nabla_\mu\nabla_\nu H
\nonumber
\\
&&\qquad\qquad\qquad{}
-g_{\mu\nu}(\nabla^2-m^2+\tfrac{R}{d})H
+g_{\mu\nu}\nabla^\alpha\nabla^\beta H_{\alpha\beta}
=0.
\label{EM}
\end{eqnarray}
If we consider equation (\ref{Det}) for equations (\ref{EM}) then we
find that it fulfils identically. Therefore one should replace
equations (\ref{EM}) by another equivalent system of equations with
constraints on initial data. It can be done by the same method as in
\cite{Velo} and we will not repeat all the steps and proofs. The
system of equations equivalent to (\ref{EM}) are
\begin{eqnarray}
E_{\mu\nu}+\nabla_\mu C_\nu+\nabla_\nu C_\mu+\nabla_\mu\nabla_\nu D=0,
\label{EM-2}
\end{eqnarray}
where
\begin{eqnarray}
C_\nu&=&-\frac{1}{m^2}\nabla^\mu E_{\mu\nu}
=\nabla^\mu H_{\mu\nu}-\nabla_\nu H,
\\
D&=&\frac{d-2}{m^2[(d-1)m^2+R]}
\Bigl(\nabla^\mu\nabla^\nu E_{\mu\nu}+\frac{m^2}{d-2}E^\mu_\mu\Bigr)
=H
\end{eqnarray}
and it is supplemented by the constraints at an initial time (say $t=0$)
\begin{eqnarray}
E_{\mu0}|_{t=0}=0,
\qquad
C_\nu|_{t=0}=0,
\qquad
D|_{t=0}=0,
\qquad
(\nabla_0D)|_{t=0}=0.
\end{eqnarray}
Then one can find that the kinetic part of (\ref{EM-2}) has the form
\begin{eqnarray}
\nabla^2H_{\mu\nu}+g_{\mu\nu}(\nabla^\alpha\nabla^\beta
H_{\alpha\beta}-\nabla^2H)+\ldots=0
\end{eqnarray}
and coincides at any point $x_0$ with its analog in the flat space
if we choose locally around $x_0$ the Riemann normal coordinates.
Thus like in the flat case, the equations for spin-2 field
followed from Lagrangian (\ref{L2-FP}) are hyperbolic and causal for
the background under consideration.

\section{Summary}\label{summary}
We have studied a new aspect of BRST approach to Lagrangian
construction for higher spin fields. The approach efficiently works
if we require that BRST operator is nilpotent only in weak sense,
i.e. $Q_s^2|\Lambda_s\rangle=0$ for some spin-s but in general as
the operator $Q_s^2\neq0$. In all previous applications of BRST
approach to higher spin theory the operator equality $Q^2=0$ was
valid only in constant curvature spaces for any spin, even for
spin-1 field. We have shown that nilpotency of BRST charge in weak
sense leads to standard spin-1 field Lagrangian in arbitrary Riemann
spacetime and to consistent spin-2 field Lagrangian in Einstein
spacetime. As a result we resolved an old enough puzzle that BRST
approach to field theory with arbitrary spin is unable to reproduce
the known Lagrangian for spin-1 field in curved spacetime.

As usual in BRST approach, spin-1 and spin-2 Lagrangians are
obtained in gauge invariant form with suitable St\"uckelberg
auxiliary fields. In particular, we see that the Lagrangian found in
\cite{Zinoviev} for constant curvature space is consistent in more
general spaces with nontrivial Weyl tensor. Also we extended the
notion of partial masslessness, formulated in \cite {Deser} in
constant curvature space, for spin-2 field in above background
with non-zero Weyl tensor. And at last, we proved that the
Lagrangian under consideration describes a hyperbolic and causal
propagation of spin-2 field in the above spacetime.

\section*{Acknowledgements}
The authors are grateful to Yu.~Zinoviev and M.~Tsulaia for
discussions and A.~Waldron for comments. The work is partially
supported by the grant for LRSS, project No.\ 3558.2010.2, the RFBR
grant, project No.\ 09-02-00078 and the RFBR-Ukraine grant, project
No.\ 10-02-90446. The work of I.L.B. and P.M.L. is also partially
supported by the RFBR-DFG grant, project No.\ 09-02-91349 and the
DFG grant, project No. 436 RUS 113/669/0-4.

\end{document}